\documentclass[svgnames]{aastex631}

\usepackage{graphicx}	
\usepackage{amsmath}	
\usepackage{amssymb}    

\usepackage{booktabs}
\usepackage[flushleft]{threeparttable}

\newcommand{\stagger}{\texttt{Stagger}}
\newcommand{\numax}{$\nu_{\max}$}

\begin{document}

\title{Does the \numax{} scaling relation depend on metallicity? Insights from 3D convection simulations}

\correspondingauthor{Yixiao Zhou}
\email{yixiao.zhou@qq.com}

\author{Yixiao Zhou}
\author{J{\o}rgen Christensen-Dalsgaard}
\affiliation{Stellar Astrophysics Centre, Department of Physics and Astronomy, Aarhus University, Ny Munkegade 120, DK-8000 Aarhus C, Denmark}
\author{Martin Asplund}
\affiliation{Australian Academy of Science, Box 783, Canberra, ACT 2601, Australia}
\author{Yaguang Li}
\affiliation{Institute for Astronomy, University of Hawaii, 2680 Woodlawn Drive, Honolulu, HI 96822, USA}
\author{Regner Trampedach}
\affiliation{Space Science Institute, 4765 Walnut Street, Boulder, CO 80301, USA}
\author{Yuan-Sen Ting}
\affiliation{Research School of Astronomy and Astrophysics, Australian National University, Canberra, ACT 2611, Australia}
\affiliation{Research School of Computer Science, Australian National University, Acton ACT 2601, Australia}
\affiliation{Department of Astronomy, The Ohio State University, Columbus, OH 43210, USA}
\author{Jakob L. R{\o}rsted} \altaffiliation{Formerly Jakob R{\o}rsted Mosumgaard}
\affiliation{Stellar Astrophysics Centre, Department of Physics and Astronomy, Aarhus University, Ny Munkegade 120, DK-8000 Aarhus C, Denmark}
\affiliation{Aarhus Space Centre (SpaCe), Department of Physics and Astronomy, Aarhus University, Denmark}



\begin{abstract}

Solar-like oscillations have been detected in thousands of stars thanks to modern space missions. These oscillations have been used to measure stellar masses and ages, which have been widely applied in Galactic archaeology. One of the pillars of such applications is the \numax{} scaling relation: the frequency of maximum power \numax{}, assumed to be proportional to the acoustic cut-off frequency, $\nu_{\rm ac}$, scales with effective temperature and surface gravity. However, the theoretical basis of the \numax{} scaling relation is uncertain, and there is an ongoing debate about whether it can be applied to metal-poor stars. 
We investigate the metallicity dependence of the \numax{} scaling relation by carrying out 3D near-surface convection simulations for solar-type stars with [Fe/H] between -3 and 0.5 dex. Firstly, we found a negative correlation between $\nu_{\rm ac}$ and metallicity from the 3D models. This is in tension with the positive correlation identified by studies using 1D models. Secondly, we estimated theoretical \numax{} values using velocity amplitudes determined from first principles, by quantifying the mode excitation and damping rates with methods validated in our previous works. We found that at solar effective temperature and surface gravity, \numax{} does not show correlation with metallicity. This study opens an exciting prospect of testing the asteroseismic scaling relations against realistic 3D hydrodynamical stellar models.

\end{abstract}



\section{Introduction} \label{sec:intro}

Space missions such as CoRoT \citep{2008Sci...322..558M}, \textit{Kepler} \citep{2010Sci...327..977B}, and TESS \citep{2015JATIS...1a4003R} have monitored light variations for hundreds of thousands of stars, and revolutionized the study of solar-like oscillations.
Solar-like oscillations refer to those detected in Sun-like stars, subgiants, and red giants, which are stochastically excited by near-surface convection. 
Information about the frequency and amplitude of the oscillations allows us to infer stellar properties (e.g.~\citealt{2017ApJ...835..173S,2018ApJS..236...42Y}) and detailed internal structure (e.g.~\citealt{2009ApJ...699.1403B,2019ApJ...885..143B}).

Two particularly important asteroseismic observables are the large frequency separation $\Delta\nu$ and the frequency of maximum mode-power \numax{}, from which it is possible to obtain estimates of the stellar radii and masses by using the so-called asteroseismic scaling relations that are calibrated against the Sun \citep{1986ApJ...306L..37U,1991ApJ...368..599B,1995A&A...293...87K}:
\begin{align} 
\frac{\Delta\nu}{\Delta\nu_{\odot}} &\simeq \sqrt{ \frac{\bar{\rho}}{\bar{\rho}_{\odot}} } 
= \left(\frac{M}{M_{\odot}}\right)^{1/2} \left( \frac{R}{R_{\odot}}\right)^{-3/2}, 
\label{eq:Deltanu-scale} 
\\
\frac{\nu_{\max}}{\nu_{\max,\odot}} &\simeq 
\frac{\nu_{\rm ac}}{\nu_{\rm ac,\odot}} \simeq 
\left(\frac{g}{g_{\odot}} \right) 
\left(\frac{T_{\rm eff}}{T_{\rm eff, \odot}}\right)^{-1/2},
\label{eq:numax-scale}
\end{align}
where $R$ and $M$ stand for the total radius and mass of the star, and the subscript ``$\odot$'' represents properties that are associated with the Sun. The large frequency separation $\Delta\nu$, defined as the frequency difference between two adjacent radial modes $n$ with the same spherical harmonic (or angular) degree $l$, is understood to represent the mean density $\bar{\rho}$ of a star \citep{1986ApJ...306L..37U}. The frequency of maximum oscillation amplitude \numax{} has no solid theoretical basis, but is empirically related to the acoustic cut-off frequency $\nu_{\rm ac}$ below which oscillations are trapped within the star, which in turn is connected to the stellar surface gravity $g$ and effective temperature $T_{\rm eff}$ in an isothermal atmosphere \citep{1991ApJ...368..599B}. 
If both $\Delta\nu$, $\nu_{\rm max}$ and $T_{\rm eff}$ of a star are measured, Eqs.~\eqref{eq:Deltanu-scale} and \eqref{eq:numax-scale} can be applied to directly estimate the stellar radius and mass via \citep{2009ApJ...700.1589S,2010A&A...509A..77K}:
\begin{equation} \label{eq:SSR}
\begin{aligned} 
\frac{R}{R_{\odot}} &\simeq \left(\frac{\nu_{\rm max}}{\nu_{\rm max, \odot}}\right) \left( \frac{\Delta\nu}{\Delta\nu_{\odot}}\right)^{-2} 
\left(\frac{T_{\rm eff}}{T_{\rm eff, \odot}} \right)^{1/2}, 
\\
\frac{M}{M_{\odot}} &\simeq \left(\frac{\nu_{\rm max}}{\nu_{\rm max, \odot}}\right)^3 \left( \frac{\Delta\nu}{\Delta\nu_{\odot}}\right)^{-4} 
\left(\frac{T_{\rm eff}}{T_{\rm eff, \odot}} \right)^{3/2}.
\end{aligned}
\end{equation}
We refer the reader to \citet{2020FrASS...7....3H} for a thorough review of the asteroseismic scaling relation and its applications.

The asteroseismic scaling relations \eqref{eq:Deltanu-scale}, \eqref{eq:numax-scale} [equivalently \eqref{eq:SSR}] are widely used for the determination of fundamental stellar parameters due to the simplicity of this method. For example, accurate stellar radii, which can be obtained from the scaling relation, are important for characterizing exoplanet systems \citep{2013ApJ...767..127H}; asteroseismic masses has been applied to quantify the mass loss along the red-giant branch in star clusters \citep{2012MNRAS.419.2077M}; stellar ages constrained by asteroseismic radii and masses are crucial to understand the evolution of our Galaxy \citep{2018MNRAS.475.5487S}.

Given their great significance, it is necessary to examine the validity of the scaling relations. By comparing the radii based on the scaling relation \eqref{eq:SSR} against those derived from \textit{Gaia} parallaxes, \citet{2017ApJ...844..102H} found good agreement (at a 5\% level) for stars with radii between $0.8$ and $10$  $R_{\odot}$ (see also \citealt{2018MNRAS.476.1931S}). 
\citet{2016ApJ...832..121G} and \citet{2018MNRAS.476.3729B} tested the asteroseismic mass for three oscillating red giants in eclipsing binaries. It turns out that the scaling relation \eqref{eq:SSR} overestimates masses by about 15\% compared to dynamical masses.

Tests on individual scaling relations were also carried out. \citet{2011ApJ...743..161W}, among other works \citep{2016MNRAS.460.4277G,2016ApJ...822...15S,2017MNRAS.467.1433R,Serenelli++2017}, calculated $\Delta\nu$ and $\bar{\rho}$ from stellar structural models. They highlighted deviations from the $\Delta\nu$ scaling given in \eqref{eq:Deltanu-scale} and introduced a correction factor $f_{\Delta\nu}$.
By applying this $f_{\Delta\nu}$-corrected scaling relation, it was found that the radii align more closely with \textit{Gaia} radii within 2\% \citep{2019ApJ...885..166Z}. Moreover, the masses now agree with dynamical masses within observational uncertainties \citep{2018MNRAS.476.3729B,2022A&A...668A..82B}.
On the other hand, based on the mixing length theory of convection and results from non-adiabatic calculations of mode excitation and damping, \citet{2011A&A...530A.142B,2013ASPC..479...61B} suggested that the \numax{} scaling relation should depend on Mach number near the stellar surface. \citet{2015MNRAS.451.3011C} used model-derived stellar properties to rule out departures from the \numax{} scaling \eqref{eq:numax-scale} above 1.5\%. In a separate effort, \citet{Li++2021} examined the intrinsic scatter of the scaling relations due to hidden variables, which was shown to be nearly negligible.

However, \citet{2014ApJ...785L..28E} studied nine metal-poor halo and thick-disk red giants and found their asteroseismic masses obtained from \eqref{eq:SSR} are systematically higher than expectations by about 15\%. The overestimation of stellar masses remains even if the corrected $\Delta\nu$ scaling is adopted, therefore motivating the examination of the \numax{} scaling relation in the metal-poor regime. \citet{2017ApJ...843...11V} suggested the \numax{} scaling relation should depend on mean molecular weight (hence metallicity), based on the relationship between acoustic cut-off frequency and fundamental stellar parameters. \citet{Bellinger++2019} and \citet{2022ApJ...927..167L} compared stellar properties estimated from the scaling relations with those obtained from individual frequency modeling. They clearly showed that a metallicity term in the scaling relations is needed in order to reach better agreements.
Yet, as pointed out by \citet{2023arXiv230410654S}, the overestimation of masses for metal-poor stars and the trend with metallicity might arise from offsets in the effective temperature scale rather than errors in the scaling relation.

In this study, we aim to answer whether the \numax{} scaling relation depends on metallicity from 3D hydrodynamical simulations. \citet{2019ApJ...880...13Z,2020MNRAS.495.4904Z} demonstrated that 3D simulations of stellar near-surface convection can be used to quantify the excitation, damping, and oscillation amplitude of pressure modes (p-modes) in a self-consistent, parameter-free manner. Here we follow the theoretical formulation and numerical technique developed in previous works to directly estimate \numax{} (Sect.~\ref{sec:amplitude}) based on six 3D models with almost identical surface temperature and gravity but spanning a wide range of metallicity. 
The correlation between theoretical \numax{} and metallicity of the 3D model reflects the role of metallicity in the \numax{} scaling relation, independent of asteroseismic observations.
We also compute the important intermediate component of the \numax{} scaling relation, the acoustic cut-off frequency, from 3D models and investigate its relationship with metallicity (Sect.~\ref{sec:nuac}). As a first step, the current work focuses on solar effective temperature and surface gravity.

\section{Modeling} \label{sec:model}

\subsection{3D stellar atmosphere models} \label{sec:3D-model}

\begin{table}
\centering
\begin{threeparttable}
\caption{Fundamental parameters and basic properties of 3D near-surface convection simulations used in this work. Except for the solar model, model names are constructed from the effective temperature (labeled as ``\texttt{t}''), surface gravity (``\texttt{g}''), and metallicity (``\texttt{p}'' for positive while ``\texttt{m}'' for negative metallicity). Since effective temperatures derived from the radiative transfer calculations fluctuate over time, both the mean value and its standard deviation are shown. All models have the same surface gravity, which equals the nominal solar value of \citet{2016AJ....152...41P}. Numerical resolution is not uniform in the vertical direction, therefore both the minimum and the maximum grid spacing are given. 
Artificial driving simulations are computed with modified bottom boundary conditions. They are particularly designed for the evaluation of damping rates (cf.~Sect.~\ref{sec:amplitude}). All artificial driving simulations have identical numerical resolutions in both space and time.
\label{tb:simu-info}}
{\begin{tabular*}{\columnwidth}{@{\extracolsep{\fill}}lcccccc}
\toprule[2pt]
  Model name            & \texttt{t58g44p05}    & \texttt{solar}   & \texttt{t58g44m05}    & \texttt{t58g44m10}    & \texttt{t58g44m20}    & \texttt{t58g44m30}
  \\
\midrule[1pt]
  \multicolumn{7}{c}{Standard simulations}
  \\
\midrule[1pt]
  $T_{\rm eff}$ (K)     & $5816 \pm 17$ & $5772 \pm 16$ & $5794 \pm 9$  & $5768 \pm 9$  & $5780 \pm 9$  & $5782 \pm 6$
  \\
  $\log g$ (cgs)        & 4.438 & 4.438 & 4.438 & 4.438 & 4.438 & 4.438 
  \\
  $\rm [Fe/H]$ (dex)    & 0.5   & 0     & -0.5  & -1 (a)& -2 (a)& -3 (a)
  \\
  Numerical grids               & $240^2 \times 230$    & $240^2 \times 230$    & $240^2 \times 230$    & $240^2 \times 230$    & $240^2 \times 230$    & $240^2 \times 230$
  \\
  Time duration (hour)          & 25    & 25    & 25    & 25    & 25    & 25    
  \\
  Sampling interval (s)         & 30    & 30    & 30    & 30    & 30    & 30    
  \\
  Horizontal size (Mm)          & 8.0   & 6.0   & 8.3   & 7.5   & 6.6   & 7.2
  \\
  Vertical size (Mm)            & 4.0   & 3.6   & 3.5   & 3.4   & 3.3   & 3.4
  \\
  Vertical grid spacing (km)    & 8--49 & 7--33 & 9--31 & 8--34 & 7--37 & 8--33
  \\
\midrule[1pt]
  \multicolumn{7}{c}{Artificial driving simulations}
  \\
\midrule[1pt]
  Numerical grids               & $140^2 \times 158$    & $120^2 \times 158$    & $120^2 \times 158$    & $120^2 \times 158$    & $120^2 \times 158$    & $120^2 \times 158$
  \\
  Time duration (hour)          & 50    & 50    & 50    & 50    & 50    & 50    
  \\
  Sampling interval (s)         & 30    & 30    & 30    & 30    & 30    & 30    
  \\
  Horizontal size (Mm)          & 7.0   & 6.0   & 6.0   & 6.0   & 6.0   & 6.0
  \\
  Vertical size (Mm)            & 4.3   & 4.3   & 4.3   & 4.3   & 4.3   & 4.3
  \\
  Vertical grid spacing (km)    & 14--59& 14--59& 14--59& 14--59& 14--59& 14--59
  \\
\bottomrule[2pt]
\end{tabular*}}

    \begin{tablenotes}
      \item Note (a): $\alpha$-enhacement $\rm [\alpha/Fe] = 0.4$.
    \end{tablenotes}
    
\end{threeparttable}
\end{table}

  3D model atmospheres used in this study are computed with the \stagger{} code \citep{1995...Staggercodepaper,2018MNRAS.475.3369C,stein:StaggerCode}, a magneto-hydrodynamics (MHD) code that solves the equation of mass, momentum and energy conservation together with the equation of radiative transfer in three spatial dimensions and time. Magnetic fields can be considered by including the induction equation. 
  Partial differential equations are discretized in Cartesian grids, with scalars evaluated at the center of numerical cells while momentum vectors are staggered at cell faces and magnetic fields at edge centers.
  In this work, we ignore the effect of the magnetic field but consider the radiative transfer processes in detail. The radiative transfer equation is solved at every numerical time step for every mesh point using the long-characteristic \citet{1964CR....258.3189F} method. Spatially, we consider nine directions in the radiative transfer problem, which consists of one vertical plus eight inclined directions representing the combination of two polar angles and four azimuthal angles. The total radiative cooling (heating) rate at a given wavelength is obtained by integrating over the solid angle. The integration over the polar angle is approximated by the Radau quadrature. The numerical technique employed by the \stagger{} code is explained in detail in \citet{1995...Staggercodepaper} and \citet{2018MNRAS.475.3369C}.

  Our simulations are carried out with realistic input physics. We use a customized version of the \citet{1988ApJ...331..815M} equation of state \citep{2013ApJ...769...18T} that accounts for all ionization stages of the 17 most abundant elements in the Sun plus the hydrogen molecules. We include a comprehensive collection of continuum absorption and scattering sources as described in \citet[their Table D.1]{2010A&A...517A..49H} and \citet[Sect.~3.1]{2014MNRAS.442..805T}. Line opacities are taken from the MARCS opacity sampling data covering wavelength from 91 nm to 20 $\rm \mu m$ \citep{2008A&A...486..951G,2008PhST..133a4003P}.
  In order to reduce the computational cost, monochromatic opacities at a vast number of wavelengths are grouped into 12 ``opacity bins'' based on their approximate formation height (reflecting the strength of opacity), and wavelength \citep{1982A&A...107....1N,2013A&A...557A..26M,2018MNRAS.475.3369C,2023A&A...677A..98Z}. The organization of opacity bins is carefully adjusted such that the predicted radiative cooling rates resemble those from monochromatic calculations.

  To study the effect of metallicity on the \numax{} scaling relation, we consider six models with [Fe/H] ranging from 0.5 to -3, which covers all solar-like oscillating stars discovered to date. We adopt the \citet{2009ARA&A..47..481A} solar metal mixture and apply an abundance enhancement for $\alpha$-elements of 0.4 dex for metal-poor models with $\rm [Fe/H] \leq -1$. 
  In our models, the helium mass fraction $Y$ increases slightly with decreasing [Fe/H], from $Y=0.2422$ at $\rm [Fe/H] = 0.5$ to 0.2526 at $\rm [Fe/H] = -3$. This trend is at odds with the picture of helium enrichment in the Universe \citep{2003Sci...299.1552J}. Future \stagger{} model atmospheres constructed with updated microphysics need to take into account the consensus that $Y$ increases with metal mass fraction $Z$.
  All models are constructed with solar surface gravity and their effective temperatures are close to the solar value. Our 3D model atmospheres cover a small part of the star. Horizontally, the simulation domain has a square shape. The horizontal area is small compared with the surface area of the star but is large enough to enclose at least ten granules at any time of the simulation \citep{2013A&A...557A..26M}. Vertically, the bottom boundary of the simulation is located in the near-surface convection zone, about 2.5--3 Mm below the optical surface. The simulation domain straddles the stellar photosphere and extends up to Rosseland optical depth $\log\tau_{\rm Ross} \sim -7$. 
  Boundaries are periodic in the horizontal directions. The default bottom boundary condition of our simulations is that all inflows (fluid moving toward the surface) have identical, invariant thermal (gas plus radiation) pressure and entropy, whereas thermal pressure of outflows are adjusted to fulfill the equation of motion. 
  For all models constructed with the default boundary conditions, there are 240 grid points in each horizontal direction and 230 points along the vertical. The spacing between two adjacent grid points is constant horizontally, whereas the numerical resolution is not uniform in the vertical direction: the highest vertical resolution is applied near the optical surface in order to sufficiently capture the transition from optically thick to thin regime. 
  Our 3D simulations span a long time sequence for better resolution in the frequency domain. All six models for computing acoustic cutoff frequencies (Sect.~\ref{sec:nuac}) and mode excitation rates (Sect.~\ref{sec:amplitude}) have the same time duration of 25 hours stellar time with one simulation snapshot saved every 30 seconds, which translates to a frequency resolution of $\approx 11 \; \rm \mu Hz$.
  Their basic properties are summarized in Table \ref{tb:simu-info}. We note that the solar model is nearly identical to that used by \citet{2019ApJ...880...13Z} but extends slightly longer in time, while the five models with non-solar metallicities originate from the \stagger{}-grid \citep{2013A&A...557A..26M} with small adjustments in surface gravity to agree with the value used in our solar model.

Computing mode damping rates from near-surface convection simulations for a number of frequencies is challenging. Following \citet{2019ApJ...880...13Z,2020MNRAS.495.4904Z}, we carried out numerical experiments where we artificially excite radial oscillations by perturbing the bottom boundary at a fixed frequency. 
Specifically, we add a small sinusoidal term on top of the constant thermal pressure of the incoming flows while keeping the entropy of inflows constant (in first order) at the bottom boundary:
\begin{equation} \label{eq:bbc}
    \begin{aligned}
    P_{\rm bot} &= P_{\rm bot,0} \left[1 + \epsilon\sin \omega_{\rm d} t \right],
    \\
    s_{\rm bot} &= s_{\rm bot,0} + \mathcal{O}(\epsilon^2),
    \end{aligned} 
\end{equation}
where $P_{\rm bot,0}$ and $s_{\rm bot,0}$ are constant bottom thermal pressure and entropy per mass used in standard simulations. The term $\epsilon$ is a small, dimensionless number that governs the amplitude of perturbation, $\omega_{\rm d}$ is the angular frequency of artificial driving, and $t$ stands for time. The perturbation at the bottom boundary will lead to coherent radial oscillations throughout the simulation domain. Detailed description and validation of this numerical technique has been presented by \citet{2020MNRAS.495.4904Z}.

  These artificial driving simulations are specifically designed for evaluating mode damping rates. For the purpose of controlling variables, all artificial driving simulations share the same numerical resolution (the distance between two adjacent grid points) in all dimensions. The solar and metal-poor models cover $6 \times 6 \; \rm Mm^2$ area horizontally while the horizontal size of model \texttt{t58g44p05} is $7 \times 7 \; \rm Mm^2$ to ensure that at least ten granules are always included in the simulation domain\footnote{At given $T_{\rm eff}$ and $\log g$, the size of granulation increases with metallicity \citep{2007A&A...469..687C,2014arXiv1405.7628M}.}. The metal-rich model therefore has more mesh points in horizontal directions (Table \ref{tb:simu-info}).
  The artificial driving simulations span 50 hours of stellar time. Given the size of the simulation domain, we crudely estimate the linewidths of the fundamental (frequency about 2 mHz) and first overtone simulation mode (frequency about 3 mHz, not to be confused with the driven oscillation) are in the order of $1-10 \; \rm \mu Hz$ and $10-50 \; \rm \mu Hz$, respectively (see also \citealt{2019A&A...625A..20B}). These linewidths correspond to $e$-folding lifetimes of $8.8$--$88$ hours and $1.76$--$8.8$ hours, respectively\footnote{Due to the limited size of the simulation, the mode mass (inertia) of the simulation modes are smaller than stellar p-modes, therefore simulation modes have relatively larger linewidths and shorter lifetimes.}. Time sequences of the artificial driving simulations thus span at least half of the lifetime of the fundamental simulation mode and several lifetimes for the first overtone mode.

\subsection{Stellar structural models} \label{sec:1D-model}

  Models of stellar structure are necessary for the calculation of mode eigenfunctions and mode masses, which are important components for evaluating excitation and damping rates (Sect.~\ref{sec:amplitude}). In the solar case, we use the standard solar model of \citet[model S]{1996Sci...272.1286C}. 
  At other metallicities, stellar interior models are computed with the Garching Stellar Evolution Code (\texttt{GARSTEC}; \citealt{2008Ap&SS.316...99W}). However, the interior model is only constrained by atmosphere parameters that define 3D models, namely $T_{\rm eff}$, $\log g$, and [Fe/H]. Therefore, different combinations of input parameters such as mass and mixing length parameter $\alpha_{\rm MLT}$ might give rise to multiple models that fit the three constraints well. 
  In order to break this degeneracy, we construct stellar structural models based on a novel method developed by \citet{2018MNRAS.481L..35J} and \citet{2020MNRAS.491.1160M}, which couples the (interpolated) mean 3D stratification from the \stagger{}-grid with the 1D interior model at every step of stellar evolution. In practice, we replace the near-surface regime of the 1D interior model with the (interpolated) mean 3D model and set the outer boundary condition of the stellar evolution calculation from the mean 3D model in every timestep of stellar evolution (cf.~\citealt{2018MNRAS.481L..35J} and \citealt{2020MNRAS.491.1160M} for a detailed description of the methodology). 
  As the outer boundary of the interior model is located in the convection zone where the temperature gradient is close to adiabatic, the evolution calculation is not sensitive to $\alpha_{\rm MLT}$ (\citealt{2020MNRAS.491.1160M} Sect.~3.2). To this end, we fix $\alpha_{\rm MLT}$ to the solar-calibrated value and only adjust the stellar mass to obtain the best-fitting model at non-solar metallicities. We also note that stellar structural models generated with the 3D-1D coupling method are consistent with 3D model atmospheres used in this study in terms of mean structures near the stellar surface.

\section{The effect of metallicity on the acoustic cut-off frequency} \label{sec:nuac}

\begin{figure}
\includegraphics[width=\textwidth]{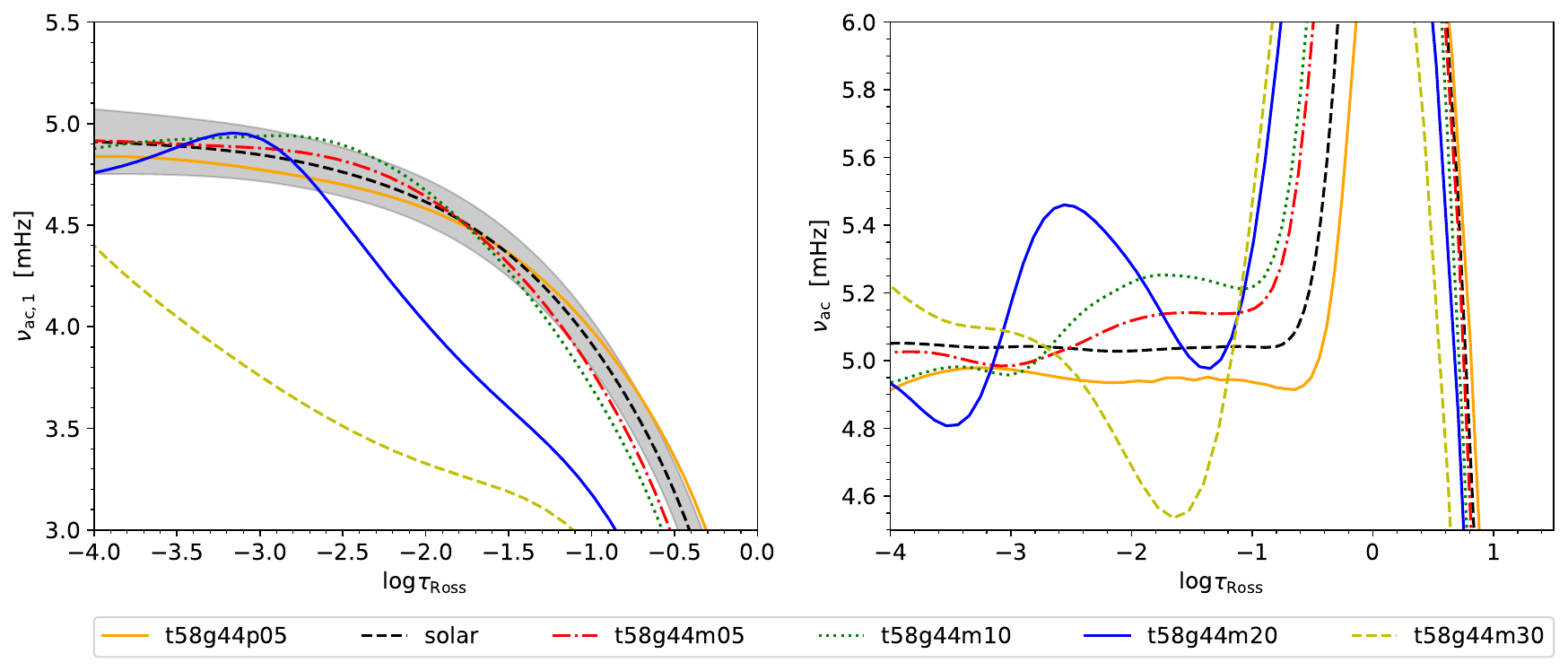}
\caption{\textit{Left panel:} First-order acoustic cut-off frequency computed via Eq.~\ref{eq:nuac1} as a function of spatially and temporally averaged Rosseland optical depth for all models included in this work. The gray shaded region outlines standard deviations of the temporal fluctuations of $\nu_{\rm ac,1}$ for the solar model. We only show the distribution of acoustic cut-off frequencies up to $\log\tau_{\rm Ross} = -4$, as our models are likely less reliable in regions higher up in the atmosphere due to the neglect of magnetic fields.
\textit{Right panel:} The distribution of acoustic cut-off frequencies computed based on the general definition (Eq.~\eqref{eq:nuac}).
\label{fig:nuac}}
\end{figure}

\begin{figure}
\centering
\includegraphics[width=0.5\textwidth]{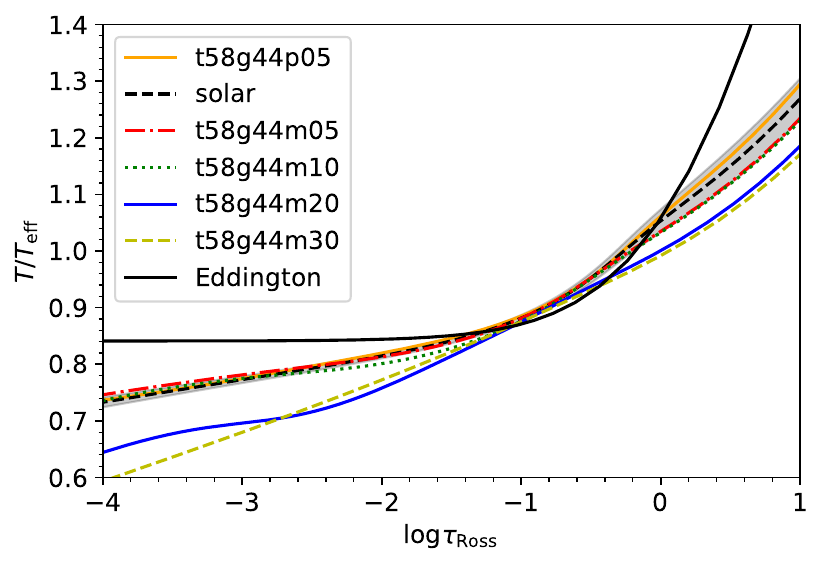}
\caption{The relationship between horizontal- and time-averaged temperature and Rosseland optical depth ($T - \tau$ relation) predicted by 3D model atmospheres at different metallicities. The Eddington $T - \tau$ relation, which is an approximation to the gray atmosphere, is shown in solid black line. 
\label{fig:Ttau}}
\end{figure}

  The acoustic cut-off frequency, which can be understood as a barrier to acoustic waves, is an important quantity that motivates the link of \numax{} to $g/\sqrt{T_{\rm eff}}$. Assuming the stellar atmosphere is isothermal, the kinetic energy of p-modes with frequency lower than $\nu_{\rm ac,1}$ will decay exponentially in the stellar atmosphere, hence trapped in the stellar interior. Higher frequency modes with $\nu > \nu_{\rm ac,1}$, also called pseudo-modes \citep{1995MNRAS.272..850R,2007MNRAS.381.1001K}, are able to propagate through the atmosphere.
  \citet{2017ApJ...843...11V} proposed that under the assumption of isothermal atmosphere and ideal gas equation of state, 
  \begin{equation} \label{eq:nuac1}
      \nu_{\rm ac,1} = \frac{c_s}{4\pi H_{\rho}} \propto g \sqrt{\frac{\mu}{T_{\rm eff}}},
  \end{equation}
  where $\mu$ represents the mean molecular weight, which originates from the ideal gas equation of state and was ignored in the derivation of the canonical \numax{} scaling relation. Eq.~\eqref{eq:nuac1} suggests that at fixed effective temperature and surface gravity, the acoustic cut-off frequency increases with metallicity (see Fig.~13 of \citealt{2017ApJ...843...11V}).
  A more general expression of the acoustic cut-off frequency was derived in \citet{1984ARA&A..22..593D}. For adiabatic oscillations whose wavelength is much less than the stellar radius, the acoustic cut-off frequency reads
  \begin{equation} \label{eq:nuac}
      \nu_{\rm ac} = \frac{c_s}{4\pi H_{\rho}} \sqrt{1 - 2\frac{dH_{\rho}}{dr}},
  \end{equation}
  which reduces to $\nu_{\rm ac,1}$ in the isothermal scenario.

  We compute acoustic cut-off frequencies in the isothermal limit (Eq.~\eqref{eq:nuac1}) as well as from the general definition (Eq.~\eqref{eq:nuac}) for all 3D models. All components in Eqs.~\eqref{eq:nuac1} and \eqref{eq:nuac} are mean quantities averaged over constant geometric depth (horizontal average) and the entire time series. The distribution of acoustic cut-off frequencies near the optical surface and in the lower stellar atmosphere is shown in Fig.~\ref{fig:nuac}.
  The violent oscillation of $\nu_{\rm ac}$ just below the optical surface (\textit{right panel} of Fig.~\ref{fig:nuac}) is associated with the large super-adiabatic temperature gradient at the top of the hydrogen ionization zone \citep{1995ApJ...440..297S}. This gives rise to a ``density inflection region'' (cf.~Fig.~1 of \citealt{2020MNRAS.491.1160M}) and a rapid change in density scale height. Because acoustic cut-off frequencies change with optical depth in the stellar atmosphere, it is difficult to determine an exact $\nu_{\rm ac}$ value from our models. Nevertheless, we can see that the maximum of $\nu_{\rm ac,1}$ and $\nu_{\rm ac}$ within $-4 \leq \log\tau_{\rm Ross} \leq -1$ anti-correlate with metallicity when $\rm -1 \leq [Fe/H] \leq 0.5$.
  This trend is in tension with the conclusion of \citet{2017ApJ...843...11V}, who suggested at given $T_{\rm eff}$ and $\log g$, acoustic cut-off frequency increases with metallicity because of the dependence on the mean molecular weight (see their Fig.~13). 
  We remark that, unlike the present work, \citet{2017ApJ...843...11V} assumed the helium enrichment law of $\Delta Y / \Delta Z = 1$ or $2$ in their calculations of $\mu$, whereas helium anti-correlates with metal mass fraction in our 3D models (Sect.~\ref{sec:model}). With this in mind, we computed mean molecular weights for all 3D models and found that $\mu$ in stellar atmospheres decreases with metallicity in the range $\rm -1 \leq [Fe/H] \leq 0.5$. Therefore, different helium mass fractions used in \citet{2017ApJ...843...11V} and this work is not the main cause of the tension in the acoustic cut-off frequency.
  The disagreement is associated with the assumption of isothermal atmosphere when deriving Eq.~\eqref{eq:nuac1}. Although isothermal is a good approximation for gray atmospheres, temperature profiles predicted by 3D models are far from isothermal and show substantial departure from the solar relation between $T$ and $T_{\rm eff}$ (Fig.~\ref{fig:Ttau}). To this end, the full expression of the acoustic cut-off frequency is more appropriate in our context because it does not rely on the assumption of an isothermal atmosphere. 
  
  On the other hand, the physical meaning of acoustic cut-off frequency becomes quite ambiguous in the case of very metal-poor models \texttt{t58g44m20} and \texttt{t58g44m30}, where their $\nu_{\rm ac}$ fluctuate considerably with optical depth. This is to a large extent due to their steep temperature gradient in the atmosphere (Fig.~\ref{fig:Ttau}). We note that temperature stratification in the optically thin layers is largely dictated by the relative strength of adiabatic cooling in the upflows and radiative heating due to the absorption of photons \citep{1999A&A...346L..17A,2007A&A...469..687C}. For the [Fe/H]$=-3$ simulation, with the weakest line opacity, the balance between radiative heating and adiabatic cooling is dominated by the latter, [Fe/H]$=-2$ is a transition case, and [Fe/H]$\ge -1$ is dominated by radiative heating. 
  In short, 3D model atmospheres suggest that in the range of $\rm -1 \leq [Fe/H] \leq 0.5$, the acoustic cut-off frequency increases with decreasing metallicity.

\section{Estimating mode excitation, damping rates, and \numax{}} \label{sec:amplitude}

\begin{figure}
\centering
\includegraphics[width=0.5\textwidth]{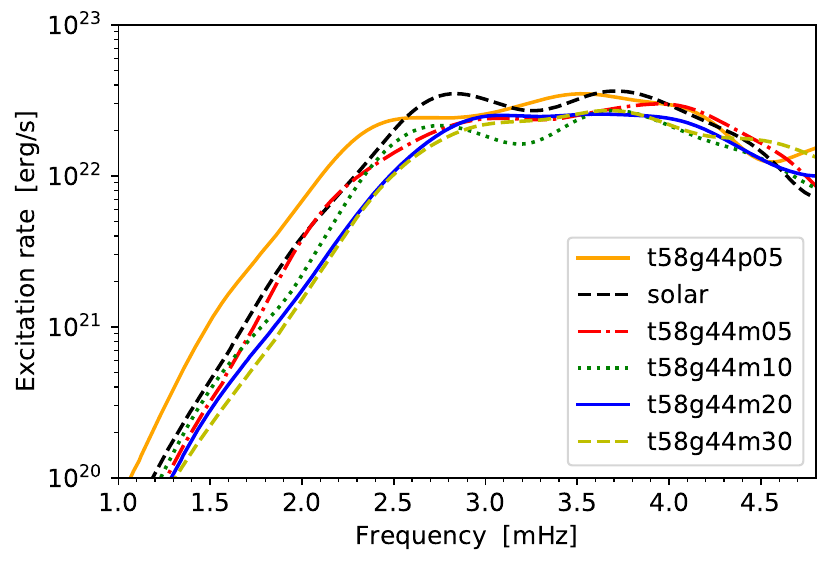}
\caption{Excitation rates of radial oscillations, $\mathcal{P}_{\rm exc}$, as a function of cyclic frequency computed via Eq.~\eqref{eq:Pexc} for six models at solar $T_{\rm eff}$ and $\log g$. All curves shown here are smoothed from the original results with Gaussian kernels with a full width at half maximum (FWHM) of 0.47 mHz.
\label{fig:Pexc}}
\end{figure}

\begin{figure}
\includegraphics[width=0.5\textwidth]{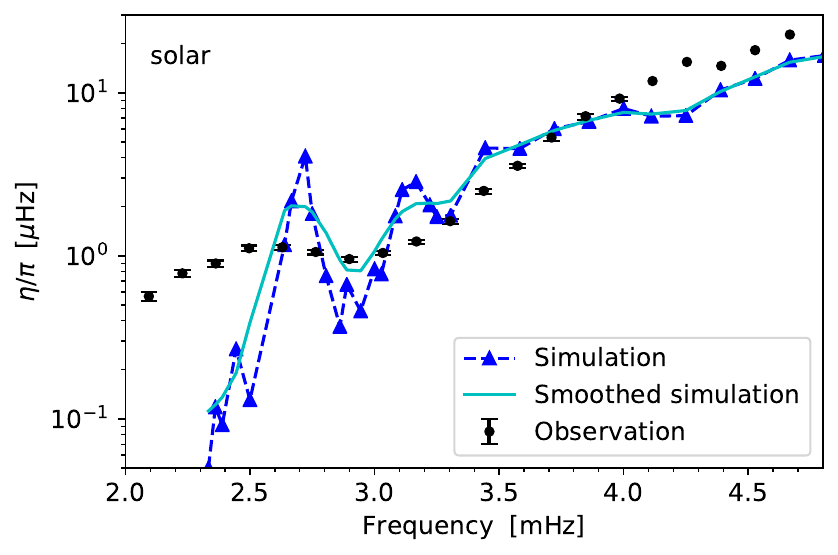}
\includegraphics[width=0.5\textwidth]{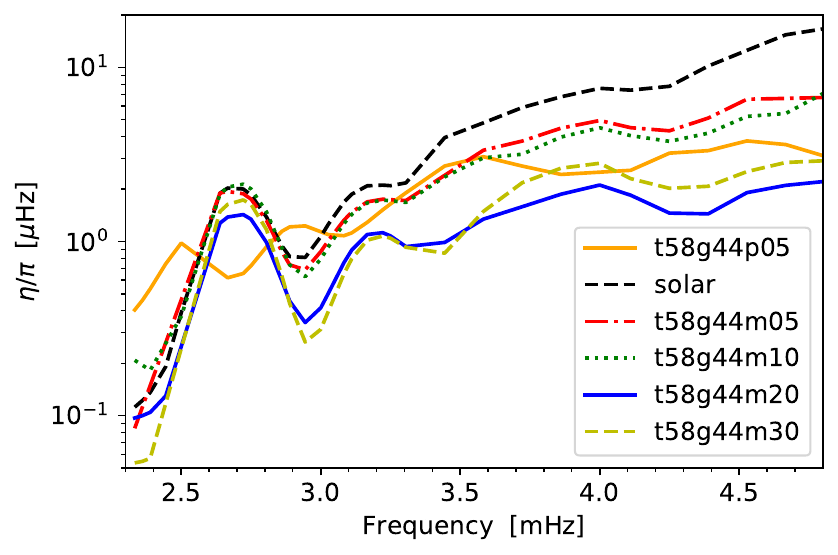}
\caption{\textit{Left panel:} Linear damping rates computed from artificial driving solar simulations at different cyclic frequencies (blue triangles) are divided by $\pi$ to compare with observed line widths from BiSON $l=0$ data (black dots). Below 4 mHz, the observational data, with uncertainties, is taken from \citet{2005MNRAS.360..859C}. BiSON line widths above 4 mHz are derived from Table 1 of \citet{1998MNRAS.298L...7C}, therefore uncertainties are not accessible. The cyan line is obtained by smoothing the raw simulation data with a Gaussian kernel with an FWHM equal to 0.2 mHz.
\textit{Right panel:} Distribution of smoothed damping rates for six models with different metallicities. An identical smoothing kernel is used for all models. The black dashed line corresponds to the cyan line in the \textit{left panel}.
\label{fig:linew}}
\end{figure}

\begin{figure}
\centering
\includegraphics[width=0.5\textwidth]{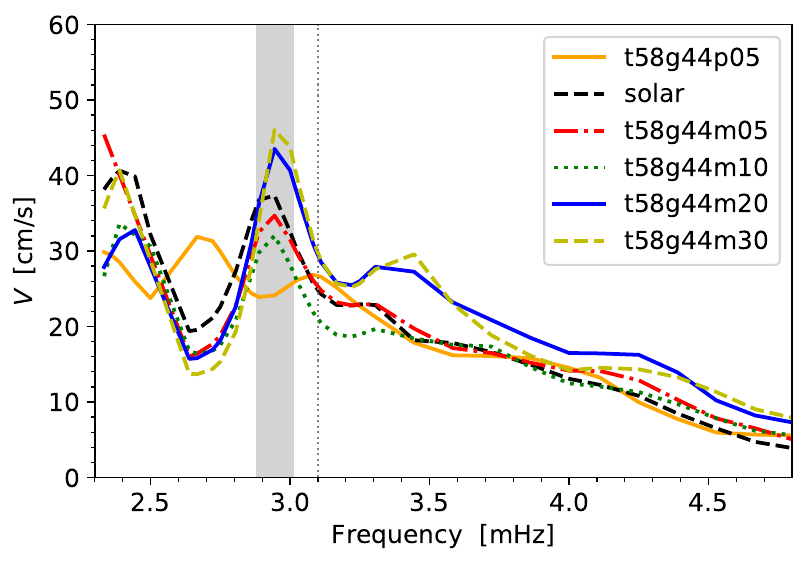}
\caption{Predicted photospheric velocity amplitudes for models with different metallicities. The range of theoretical \numax{}, $2.88 \leq \nu_{\max} \leq 3.01$ mHz, estimated for the solar and all sub-solar metallicity models is shaded in gray. The vertical dotted gray line marks the measured \numax{} for the Sun \citep{2008ApJ...682.1370K}.
\label{fig:Vamp}}
\end{figure}

  In solar-type stars, turbulent convection near the stellar surface stochastically excites numerous p-modes at different frequencies, appearing as peaks in the observed power spectrum. Observationally, oscillation amplitudes are measured by first smoothing the raw power spectrum and then subtracting the background noise (e.g.~\citealt{2005ApJ...635.1281K,2009A&A...495..979M}). The frequency corresponding to the peak of the amplitude spectrum is \numax{}. 
  However, because the extent of our simulation domain is very small compared to the whole star, only 3--4 radial modes are spontaneously excited in the simulation box\footnote{We note that high-degree non-radial modes are present in surface convection simulation, see e.g., \citet{2000ApJ...530L.139G}. However, these oscillations are only detectable for the Sun.}, and their amplitudes are not comparable to the measured amplitudes of stellar p-modes. In order to obtain realistic estimations of mode amplitudes and \numax{}, instead of analyzing the simulation modes, we quantify the energy injection (excitation) and energy dissipation (damping) rates for radial oscillations at different frequencies and derive mode amplitudes according to the energy balance between excitation and damping.

  Excitation rates of radial oscillations are computed as
\begin{equation} \label{eq:Pexc}
    \begin{aligned}
    \mathcal{P}_{\rm exc}(\omega) =&
    \frac{\omega^2 A_{\rm box}}{8 \mathcal{T}_{\rm tot} E_0} \left[ 
    \left( \int_{r_{\rm 3D \: bot}}^{r_{\rm top}} \frac{\partial \xi_r}{\partial r} 
    \mathfrak{Re}\left\lbrace \mathcal{F}[\delta \bar{P}_{\rm nad}] \right\rbrace \: dr \right)^2  \right.
    + \left. \left(\int_{r_{\rm 3D \: bot}}^{r_{\rm top}} \frac{\partial \xi_r}{\partial r} 
    \mathfrak{Im}\left\lbrace \mathcal{F}[\delta \bar{P}_{\rm nad}] \right\rbrace \: dr \right)^2 \right],
    \end{aligned}
\end{equation}
where $\mathcal{F}$ represents the Fourier transform from time to frequency domain, $\mathfrak{Re}$ and $\mathfrak{Im}$ represent real and imaginary part, respectively. The terms $\omega = 2\pi \nu$, $A_{\rm box}$, and $\mathcal{T}_{\rm tot}$ are the angular frequency, horizontal area, and total time coverage of the simulation, respectively. Numerically, the integration is carried out from the radius corresponding to the bottom boundary of the simulation domain, $r_{\rm 3D \: bot}$, to the uppermost point of the coupled stellar structure model introduced in Sect.~\ref{sec:1D-model}, $r_{\rm top}$. $\delta \bar{P}_{\rm nad}$ is the horizontally averaged non-adiabatic pressure fluctuation caused by convection. It is obtained by subtracting the adiabatic part from the total pressure fluctuation in 3D simulations, therefore including contributions from entropy fluctuation and fluctuations in turbulent pressure (Reynold stress). 
How pressure fluctuations are evaluated is detailed in \citet[their Sect.~3.2]{2019ApJ...880...13Z}.
The adiabatic eigenfunction of radial modes (also called the amplitude function) $\xi_r$ and mode kinetic energy per unit surface area
\begin{equation}
    E_0 = \frac{\omega^2}{2} 
    \int_0^{r_{\rm top}} \rho \xi_r^2(r) \left(\frac{r}{R_{\rm phot}}\right)^2 \: dr,
\end{equation}
are calculated using the Aarhus adiabatic oscillation package (\texttt{ADIPLS}, \citealt{2008Ap&SS.316..113C}) based on the coupled stellar structure model. $R_{\rm phot}$ is the photospheric radius. We note that Eq.~\eqref{eq:Pexc} is only valid for adiabatic eigenfunctions (the ``quasi-adiabatic'' approximation). The derivation of Eq.~\eqref{eq:Pexc} and the physical mechanism of p-mode excitation are explained in detail in \citet{2001ApJ...546..576N}, \citet{2001ApJ...546..585S} and \citet{2019ApJ...880...13Z}.

  Excitation rates for all models included in this study are depicted in Fig.~\ref{fig:Pexc}. Following previous works, we smooth $\mathcal{P}_{\rm exc}$ calculated from simulation data in order to eliminate noise associated with stochastic convection. Below 2.5 mHz, it is evident that mode excitation positively correlates with metallicity. However, between 2.5 and 4 mHz where $\mathcal{P}_{\rm exc}$ exhibits a plateau feature, there is no clear trend between $\mathcal{P}_{\rm exc}$ and [Fe/H].

  Damping rates of radial oscillations, which are related to the observed mode line widths $\Gamma$ through $\eta = \pi \Gamma$, are computed in frequency space according to
\begin{equation} \label{eq:eta}
    \eta(\omega) = \frac{\omega \int_{y_{\rm bot}}^{y_{\rm top}}
    \mathfrak{Im}\left\{ (\delta\bar{\rho}^{*} / \bar{\rho}_0) \delta \bar{P}_{\rm nad} \right\} dy }{4 m_{\rm mode} \vert \mathcal{V}(R_{\rm phot})\vert^2}.
\end{equation}
Here $\delta\bar{\rho}^{*}$ is the complex conjugate Fourier component of the horizontally averaged density fluctuation caused by oscillations and $\bar{\rho}_0$ is the equilibrium density obtained by averaging $\bar{\rho}$ over all simulation snapshots. $\delta\bar{\rho}^{*} / \bar{\rho}_0$ is related to the mode eigenfunction via the perturbed fluid continuity equation (cf.~\citealt{2010aste.book.....A} Chapter 3). 
In principle, the work integral in Eq.~\eqref{eq:eta} should be computed for the entire propagation cavity of the oscillation mode. Practically, the integration range is limited by the vertical coverage of the simulation, with the lower and upper bound $y_{\rm bot}$ and $y_{\rm top}$ being the geometric depth at the bottom and top simulation boundary, respectively. Nevertheless, for $n \gtrsim 15$ solar radial modes, damping processes take place around the photosphere and the super-adiabatic layers just below the optical surface according to theoretical predictions (\citealt{1992MNRAS.255..603B} Sect.~7.2 and \citealt{1996PhDT........80H} Sect.~3.4). Therefore, truncating the work integral will not affect the evaluation of damping rates for medium- and high-frequency p-modes. For all six models, our numerical results also show that contribution from layers near the bottom boundary to the work integral is negligible for oscillations with $\nu \gtrsim 2.6$ mHz, indicating that the vertical extent of our simulations is sufficient for quantifying damping rates for these higher frequency oscillations.
The denominator of Eq.~\eqref{eq:eta} is proportional to the mode kinetic energy, where $m_{\rm mode}$ is the mode mass per unit surface area and $\mathcal{V}$ is the vertical velocity amplitude. We note that $m_{\rm mode}$ is computed based on the coupled stellar structured model, whereas $\mathcal{V}$ is extracted from the 3D simulation.

  Quantifying damping rates from Eq.~\eqref{eq:eta} is challenging because it requires the knowledge of density and pressure fluctuation as well as their phase difference. Therefore, density fluctuations associated with mode eigenfunctions need to be extracted from simulations. For standard surface convection simulations, this can be done only at the frequency of the simulation mode (cf.~\citealt{2019A&A...625A..20B}), that is, p-mode oscillations naturally excited in the simulation box. Given the extent of our simulation domain, there are only 3-4 radial simulation modes from which we can compute damping rates. 
  This difficulty was overcome by carrying out multiple artificially driven simulations (see Sect.~\ref{sec:3D-model}) at different driving frequencies, which gives $\eta$ as a function of frequency\footnote{We ran about 200 artificial driving simulations in total for the evaluation of damping rates, which consumes roughly six million CPU hours at high-performance computing facilities.} (Fig.~\ref{fig:linew}).
  In the \textit{left panel}, predictions from solar simulations are compared with measured radial mode line widths from BiSON \citep{1998MNRAS.298L...7C,2005MNRAS.360..859C}. 
  In the solar case, the agreement between modeling and observation is encouraging above $\approx 2.6$ mHz. The location of the ``dip'' (i.e.~local minimum) in damping rates is also well-reproduced. Our simulations systematically underestimate damping rates below 2.6 mHz. The problem at low frequencies is a known deficiency of our approach, which can be partly attributed to the limited vertical extent of the simulation domain compared to the whole star. This cuts off significant excitation \emph{and} damping, interior of the simulations, at low frequency.
  Also, the numerically predicted damping rates demonstrate stronger fluctuations across frequencies than the measured values due to the relatively short time coverage of our simulations compared to BiSON observations, which are in the order of decades \citep{2005MNRAS.360..859C}. Nevertheless, by extending the time span of artificial driving simulations from 10 hours in \citet{2019ApJ...880...13Z} to 50 hours in the present investigation, fluctuations in $\eta$ are slightly reduced, especially in the ``plateau'' region (see Fig.~\ref{fig:linew} and \citealt{2019ApJ...880...13Z} Fig.~7). As further increasing the simulation time sequence is difficult because of practical issues such as data storage, we smooth the damping rates quantified from the simulations to make them more comparable with observations. The smoothed damping rates for all models investigated in this work are shown in the \textit{right panel} of Fig.~\ref{fig:linew}.
  Although the magnitude of $\eta$ is apparently affected by how smoothing is performed, the location of the damping rate dip, which is closely related to \numax{}, is not sensitive to the smoothing method adopted. For all models with solar and sub-solar metallicities, the depression of their damping rates is located at approximately the same frequency. For model \texttt{t58g44p05}, the local minimum of $\eta$ occurs at about 2.7 mHz, being 0.2 mHz lower than other models. We emphasize that the result of the metal-rich model is likely less reliable, as the depression in $\eta$ resides at the frequency edge where our numerical results could be trusted. 

  Once both excitation and damping rates are available, the mean kinetic velocity amplitude of the mode at the stellar surface can be quantified from the energy balance between driving and damping:
\begin{equation}
    V = \sqrt{ \frac{2\mathcal{P}_{\rm exc}}{M_{\rm mode} \eta} },
\end{equation}
where $M_{\rm mode}$ is the \emph{total} mode mass (\citealt{2010aste.book.....A} Eq.~3.140, not to be confused with the mode mass per unit surface area $m_{\rm mode}$) calculated from the coupled stellar structure model. Smoothed excitation and damping rates are used to compute the velocity amplitude, which is shown in Fig.~\ref{fig:Vamp}.
  From definition, the theoretical \numax{} is the frequency corresponding to the peak of the velocity amplitude. The thus estimated \numax{} is about 2.7 mHz for model \texttt{t58g44p05}. However, due to larger uncertainties in $\eta$ at lower frequencies, the \numax{} estimation for this model is less reliable therefore we will exclude it in subsequent discussions. 
  The frequency of maximum power predicted from the solar simulation is in reasonable agreement with the measured value ($\nu_{\max, \odot} = 3.1$ mHz, \citealt{2008ApJ...682.1370K}). The remaining discrepancy is likely due to intrinsic errors in our determination of excitation and damping rates, such as the limitation of box-in-a-star simulations or the artificial driving method. Owing to the same reason and also being aware that the smoothing kernel applied to $\mathcal{P}_{\rm exc}$ and $\eta$ affects the exact value of theoretical \numax{}, we elect to provide a range rather than exact values for \numax{} as the latter is ambiguous within our approach.
  The solar and metal-poor models all have well-defined velocity peaks located at nearly identical frequencies. Their \numax{} are estimated to be in the range of [2.88, 3.01] mHz, coinciding with the dip in damping rates.

\section{Discussion}

  Results from Sect.~\ref{sec:amplitude} lead to the conclusion that for $\rm [Fe/H] \leq 0$, the theoretical \numax{} does not demonstrate any correlation with metallicity. In order words, the \numax{} scaling relation does not depend on metallicity at solar $T_{\rm eff}$ and $\log g$.
  It is worth noting that slight differences in $T_{\rm eff}$ among our 3D models (Table \ref{tb:simu-info}) have a negligible impact on our conclusion. The mean effective temperature ranges from 5768 K for model \texttt{t58g44m10} to 5794 K for \texttt{t58g44m05}. Such differences will result in 0.23\% difference in \numax{} according to the scaling relation, translating to $\rm \approx 7 \; \mu Hz$ absolute difference in frequency. Considering the intrinsic systematics of our approach, the uncertainty associated with $T_{\rm eff}$ is minor.
  On the other hand, while the predicted velocity amplitude spectra match observations in the bulk part (\citealt{2019ApJ...880...13Z} Fig.~8), our numerical results are not sufficiently accurate to draw any conclusions about the relationship between the oscillation amplitude and [Fe/H]\footnote{A positive correlation between oscillation amplitude and [Fe/H] is expected from asteroseismic observations \citep{2018ApJS..236...42Y}. However, this trend is not seen in our numerical modeling.}.

\subsection{Why is the \numax{} scaling relation not sensitive to metallicity?}

\begin{figure}
\centering
\includegraphics[width=0.5\textwidth]{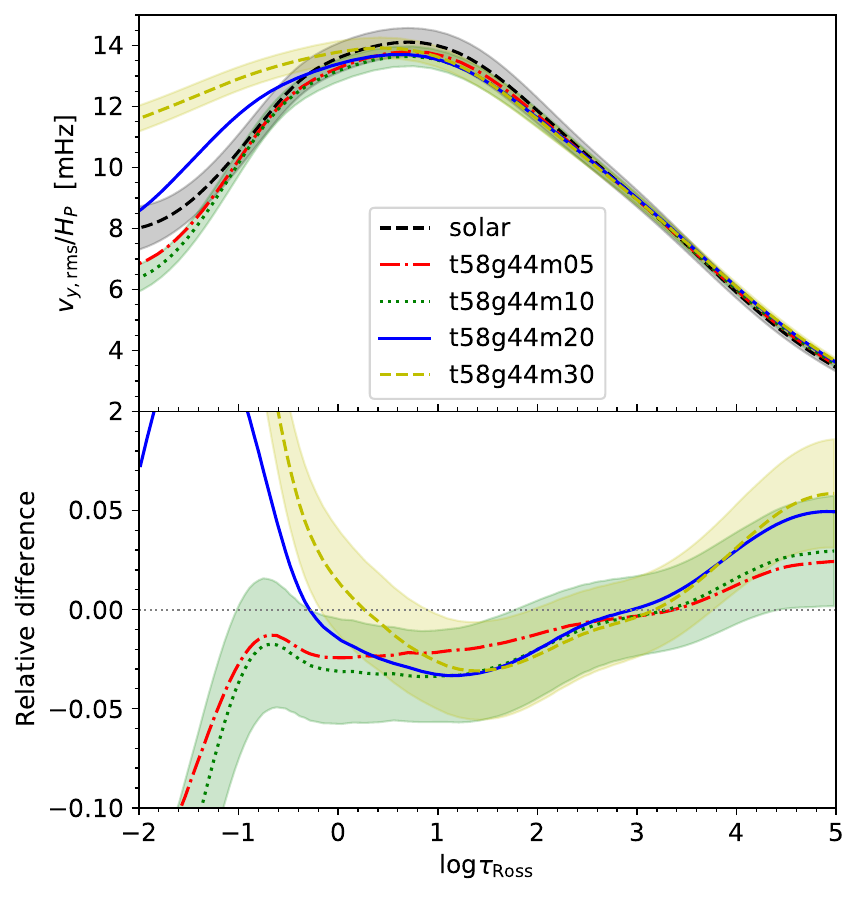}
\caption{Distribution of the convective turnover frequency in the near-surface region for the solar and metal-poor models. The turnover frequency is computed from the time-averaged rms of vertical velocity fluctuation and the horizontal- and time-averaged pressure scale height. The relative difference in $\nu_{\rm turn}$ between the metal-poor and solar model is shown in the \textit{lower panel}. Shaded areas in gray, light green and light yellow represent standard deviations of the time variation of $\nu_{\rm turn}$ (and their relative differences) for the solar model, \texttt{t58g44m10} and \texttt{t58g44m30}, respectively.
\label{fig:nuturn}}
\end{figure}

  When $\rm [Fe/H] \leq 0$, the effect of metallicity on \numax{} is less than 4.5\%. The upper limit arises from uncertainties in our numerical method rather than suggesting any correlation between \numax{} and [Fe/H]. Given the fact that stars at the same location in the HR diagram but differing in chemical composition have different stratification (e.g., $T - \tau$ relation and pressure scale height, see Fig.~\ref{fig:Ttau}) near the photosphere, this conclusion is not trivial.
  
  In Sect.~\ref{sec:amplitude}, we found that the maximum velocity amplitude is linked to the depression in damping rates, in accordance with the conclusions of \citet{1992MNRAS.255..603B} and \citet{2011A&A...530A.142B}. 
  As suggested by \citet{1992MNRAS.255..603B}, the dip in $\eta$ is caused by the negative peak (destabilizing, or exciting) in the thermal contribution to the total damping, which counteracts the convective turbulence that stabilizes (dampens) solar-like oscillations (cf.~\citealt{1992MNRAS.255..603B} Fig.~13). In 3D simulations, however, the destabilizing effect from thermal pressure fluctuations is strongest at 3.5-4 mHz \citep[Fig.~7]{2019ApJ...880...13Z}. The location of the dip around 2.95 mHz results from subtle cancelations between the positive turbulence and the negative thermal contributions to mode damping. 
  Instead of explaining the cause of the damping-rate depression with convection theory (see \citealt{2011A&A...530A.142B} for efforts in this direction), we opt to examine if the characteristic timescale (frequency) near the stellar surface depends on [Fe/H]. A frequently used timescale in the convective region is the convective turnover time, defined as the time spent for materials to travel one pressure scale height in the vertical direction \citep{2012JCoPh.231..919F}, $H_P / v_{y,\rm{rms}}$, where $v_{y,\rm{rms}}$ is the root-mean-square (rms) of vertical velocity fluctuation. Here we compare the convective turnover frequency from different models. Although our approach cannot provide a solid theoretical basis for the \numax{} scaling relation since the detailed relationship between convective turnover frequency and \numax{} is unknown, it offers useful insights into the problem in a straightforward manner.

  The distribution of convective turnover frequencies, $\nu_{\rm turn} = v_{y,\rm{rms}} / H_P$, within $-2 \leq \log\tau_{\rm Ross} \leq 5$ is depicted in Fig.~\ref{fig:nuturn} for all 3D models studied except the metal-rich one, together with the relative difference in $\nu_{\rm turn}$ with respect to the solar model. We note that layers with lower or higher optical depths are irrelevant in this context because for radial oscillations near 3 mHz, contributions to the work integral (Eq.~\eqref{eq:eta}) mostly originate from the near-surface region with $-2 \lesssim \log\tau_{\rm Ross} \lesssim 5$ (approximately from 0.2 Mm above the optical surface to 1 Mm below it in the solar case) as revealed by our artificial driving simulations.
  Fig.~\ref{fig:nuturn} shows that models with different metallicity have similar turnover frequencies near the stellar surface, with fractional differences less than 5\% when $\tau_{\rm Ross} > 0.1$. As indicated in \citet{2013A&A...557A..26M}, metal-poor stars have smaller pressure scale height and granule size. Their convective velocity is also smaller than that predicted by the solar-metallicity model in the super-adiabatic region. Here we demonstrate that these trends with [Fe/H] nearly cancel out when considering frequencies or timescales. The lack of metallicity dependence for \numax{} is thus reinforced by the insensitivity of convective turnover frequency to [Fe/H].
  Furthermore, it is worth mentioning that \citet{2022MNRAS.514.1741R} quantified the characteristic granulation timescale for four metallicities at solar $T_{\rm eff}$ and $\log g$ and found less than 4\% relative difference across $\rm -1 \leq [Fe/H] \leq 0.5$ (cf.~\citealt{2022MNRAS.514.1741R} Table 1).
  To summarize, frequencies that characterize the near-surface region, including \numax{}, are insensitive to metallicity in general, at least at solar surface temperature and gravity.

\subsection{The $\nu_{\max} - \nu_{\rm ac} - g / \sqrt{T_{\rm eff}}$ relation from the perspective of surface convection simulations}

\begin{figure}
\centering
\includegraphics[width=0.5\textwidth]{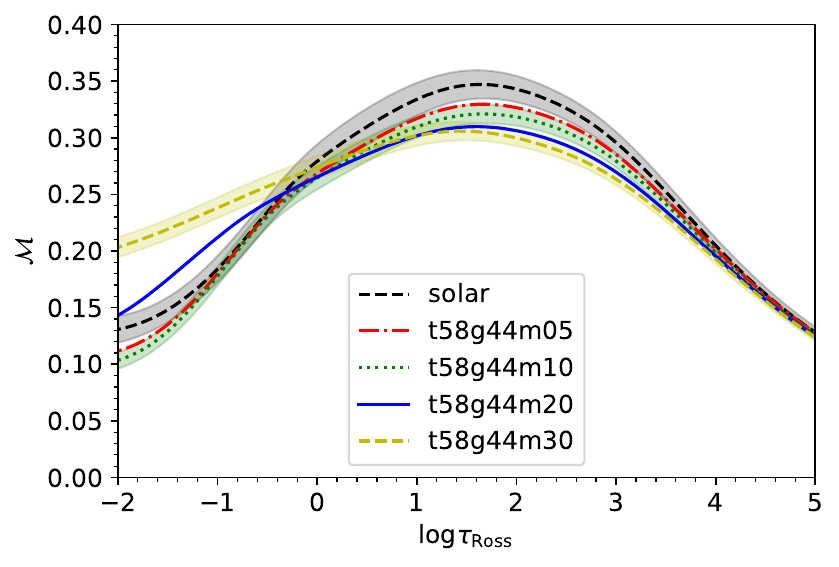}
\caption{Mach number, computed as the ratio between the rms vertical velocity fluctuation and the spatial and temporal mean of sound speed, as a function of Rosseland optical depth for five models with different metallicities. Standard deviations of the time fluctuation of Mach number are shaded for the solar model, model \texttt{t58g44m10} and \texttt{t58g44m30}.
\label{fig:Mach}}
\end{figure}

  In the original derivation of the \numax{} scaling relation \citep{1991ApJ...368..599B}, \numax{} was related to fundamental stellar parameters via the acoustic cut-off frequency. Results from 3D simulations indicate that \numax{} is insensitive to metallicity, whereas a clear anti-correlation between $\nu_{\rm ac}$ and [Fe/H] is seen within $\rm -1 \leq [Fe/H] \leq 0.5$ (Fig.~\ref{fig:nuac}). Putting these in the context of the $\nu_{\max} - \nu_{\rm ac} - g / \sqrt{T_{\rm eff}}$ relation, it is then implied that the $\nu_{\max} - \nu_{\rm ac}$ relation depends on [Fe/H] in an opposite way as the $\nu_{\rm ac} - g / \sqrt{T_{\rm eff}}$ one such that effects of metallicity cancels out for the scaling relation overall. 
  
  To assess whether this implication is reasonable, we build our analysis on the detailed physical explanation of the $\nu_{\max} - \nu_{\rm ac}$ relation proposed by \citet{2011A&A...530A.142B,2013ASPC..479...61B}, which is derived from the mixing length theory of convection, meanwhile drawing on results from non-adiabatic calculations of stellar oscillations. \citet{2011A&A...530A.142B} suggested that \numax{} is inversely proportional to the thermal timescale in the super-adiabatic region, which in turn relates to $\nu_{\rm ac}$ through thermodynamic derivatives, the Mach number, and the mixing length parameter \citep[Eq.~(15)]{2011A&A...530A.142B}. Moreover, most of the dispersion in the $\nu_{\max} - \nu_{\rm ac}$ relation is due to the Mach number. With inputs from 3D simulations, \citet{2013ASPC..479...61B} proposed that
\begin{equation} \label{eq:numax-nuac}
    \nu_{\max} \propto \mathcal{M}^{2.78} \nu_{\rm ac}.
\end{equation}
The distribution of Mach number, $\mathcal{M} = v_{y,{\rm rms}} / c_s$, in the near-surface layers predicted from our simulations is shown in Fig.~\ref{fig:Mach}.
It is evident that below the optical surface, the Mach number decreases with metallicity. Relation \eqref{eq:numax-nuac} therefore suggests a positive correlation between \numax{} and [Fe/H] at fixed $\nu_{\rm ac}$, qualitatively in line with implications from our numerical results. 
Another prominent feature of $\mathcal{M}$ is that it varies significantly in the region of interest. In combination with the fact that $\nu_{\rm ac}$ is not a constant in stellar atmosphere, these factors make quantitative interpretation of the $\nu_{\max} - \nu_{\rm ac}$ scaling \eqref{eq:numax-nuac} difficult. Here we crudely estimate the effect of metallicity on the $\nu_{\max} - \nu_{\rm ac}$ relation by arbitrarily focusing on a single depth $\tau_{\rm Ross} = 10$. At this location, the Mach number of the solar model is 1.08 times larger than model \texttt{t58g44m10}. Inserting this ratio into relation \eqref{eq:numax-nuac} gives a factor of 1.24 deviation of the $\nu_{\max} - \nu_{\rm ac}$ scaling from $\rm [Fe/H]=0$ to $-1$. Considering the difference in $\nu_{\rm ac}$ between the solar model and \texttt{t58g44m10} is roughly 3\%, the relationship \eqref{eq:numax-nuac} appears to be far too sensitive to metallicity from the perspective of our 3D simulations. 

  In short, the $\nu_{\max} - \nu_{\rm ac}$ relation proposed by \citet{2011A&A...530A.142B,2013ASPC..479...61B} suggests a positive correlation between \numax{} and [Fe/H] at given $\nu_{\rm ac}$, which qualitatively agrees with what implied from our numerical results. However, it is challenging to examine the relationship \eqref{eq:numax-nuac} quantitatively because both the Mach number and acoustic cut-off frequency vary within the near-surface region. This illustrates the complexity of the $\nu_{\max} - \nu_{\rm ac} - g / \sqrt{T_{\rm eff}}$ relation from another angle, that \numax{} is a constant, whereas the excitation and damping processes that determine its value primarily occur in the near-surface, super-adiabatic region where physical quantities vary considerably in the radial direction.

\subsection{Influence of magnetic fields on 3D surface convection simulations}

  In this study, we ignore the effect of the magnetic field. Magnetic fields can be included in the simulation by imposing a fixed field strength and orientation at the boundary of the simulation domain (e.g.~vertical magnetic fields as adopted in \citealt{2005MmSAI..76..842V} and \citealt{2023A&A...679A..65L}) or setting up an initial seed field with zero net flux and a small field strength that allow it to grow and converge to a natural state of the simulation (small-scale dynamo [SSD], e.g.~\citealt{2015A&A...578A..54T} and \citealt{2022A&A...663A.166B}). In MHD simulations of stellar surface convection, magnetic fields tend to concentrate at intergranular lanes. The average size of granules is smaller in simulations with a strong field strength \citep{2005MmSAI..76..842V,2008ApJ...684L..51J}. For the solar case, MHD models predict flatter temperature gradients near the photosphere compared with hydrodynamical models, hence have an impact on the modeled center-to-limb variations \citep{2013A&A...554A.118P,2023A&A...679A..65L}. Also, including magnetic fields will slightly reduce the pressure scale height at the optical surface. 
  Dynamically, magnetic fields inhibit surface convection, resulting in smaller convective velocities and velocity fluctuations for solar-type stars \citep{2022A&A...663A.166B}. The amplitude of simulation modes decreases with increasing field strength, indicating magnetic fields suppress p-mode oscillations in 3D simulations \citep{2008ApJ...684L..51J,2011SoPh..268..283K}. This trend is in line with the effect of solar and stellar activity on oscillation amplitude as observed by BiSON and \textit{Kepler} \citep{2000MNRAS.313...32C,2011ApJ...732L...5C}. Furthermore, helioseismic observations uncovered a positive correlation between solar activity and measured \numax{}, which shifts approximately 25 $\mu$Hz between low and high activity \citep{2020MNRAS.493L..49H}. Since the effect of magnetic fields on \numax{} is much smaller than the intrinsic systematics of \numax{} estimated from numerical simulations (Sect.~\ref{sec:amplitude}), our neglect of magnet fields will unlikely affect the result of this work.

\section{Conclusions}

  In this paper, we investigated whether the frequency of maximum power depends on metallicity at solar effective temperature and surface gravity by quantifying velocity amplitudes for radial modes based on ab initio simulations of near-surface convection covering a wide metallicity range. Velocity amplitudes were determined from the balance between energy supply and dissipation, both of which were calculated from first principles following methods described in \citet{2019ApJ...880...13Z,2020MNRAS.495.4904Z}. Our formulation and numerical approach enable purely theoretical predictions of \numax{} independent from observational data. 

  The main conclusion of this work is that when $\rm [Fe/H] \leq 0$, \numax{} does not demonstrate any trend with [Fe/H]. Furthermore, the convective turnover frequencies in the super-adiabatic layers just below the optical surface are nearly invariant across [Fe/H] as both the length scale and convective velocity are smaller in metal-poor stars, implying that for solar-type stars, frequencies characterizing the near-surface region are not sensitive to [Fe/H] in general. 
  We also calculated the acoustic cut-off frequency for all 3D models for better insights into the $\nu_{\max} - \nu_{\rm ac} - g/\sqrt{T_{\rm eff}}$ scaling. In the metallicity range $\rm -1 \leq [Fe/H] \leq 0.5$, an anti-correlation between $\nu_{\rm ac}$ and [Fe/H] is unambiguously seen, which is in tension with the conclusion of \citet{2017ApJ...843...11V}. The disagreement is attributed to the assumption of isothermal atmosphere in the derivation of the $\nu_{\rm ac} - g/\sqrt{T_{\rm eff}}$ scaling.
  On the other hand, \numax{} and $\nu_{\rm ac}$ predicted from 3D simulations suggest a positive correlation between \numax{} and [Fe/H] at fixed $\nu_{\rm ac}$. This is qualitatively in line with the $\nu_{\max} - \nu_{\rm ac}$ scaling \eqref{eq:numax-nuac} proposed by \citet{2013ASPC..479...61B}. Nevertheless, relation \eqref{eq:numax-nuac} is likely to be over-sensitive to metallicity owing to its strong nonlinear dependence on Mach number.

  For solar-type stars, the effect of metallicity on the \numax{} scaling relation is estimated to be less than 4.5\%. The upper limit here represents the intrinsic uncertainties of our numerical approach rather than indicating any possible correlation between \numax{} and [Fe/H]. 
  It is worth noting that the exact value of observed \numax{} is affected by the instrument, the data analysis process, and by the stellar activity cycle to a lesser extent \citep{2020MNRAS.493L..49H}. For example, the measured solar \numax{} ranges from 3060 (measured by the VIRGO instrument aboard the Solar and Heliospheric Observatory, \citealt{2013A&A...556A..59H}) to 3229 $\rm \mu Hz$ (measured by GOLF, \citealt{2020ASSP...57..327B}). Unless homogeneous analyses are carried out for both the Sun and other stars, the systematics in \numax{} associated with observations is comparable to the upper limit of the metallicity dependence suggested by this work. To this end, we argue that the original \numax{} scaling relation \eqref{eq:numax-scale} can be applied to metal-poor solar-type stars with confidence.

  The present analysis is confined to the parameter space of the Sun. In the next step, it will be meaningful to investigate the effect of metallicity on the \numax{} scaling relation for solar-like oscillators across the HR diagram, especially for red-giant stars. Thousands of oscillating giants from solar metallicity down to $\rm [Fe/H] \simeq -2.5$ were detected by \textit{Kepler} and TESS \citep{2018ApJS..236...42Y,2023arXiv230410654S}. Their masses and radii inferred from the asteroseismic scaling relation have important applications in Galactic archaeology and the study of globular clusters (e.g.~\citealt{2016MNRAS.455..987C,2016MNRAS.461..760M,2022A&A...662L...7T,2023MNRAS.tmp.3434H}).

  Meanwhile, we are cautious about the limitations of the adopted numerical approach. The artificial driving simulations failed to predict reasonable damping rates at low frequencies because of their small spatial and temporal coverage. The same shortcoming gives rise to fluctuations of $\eta$ across frequencies, making the final velocity amplitude affected by how smoothing is performed. These deficiencies prevent us from investigating the relationship between oscillation amplitude and metallicity, where a positive correlation was concluded from asteroseismic observations \citep{2018ApJS..236...42Y}, or studying how the width (FWHM) of the envelope of oscillation power excess changes with [Fe/H]. Resolving these problems from a theoretical angle would require accurate oscillation amplitudes for a broad range of frequencies predicted from more advanced simulations that cover the whole stellar convective envelope (e.g., \citealt{2019SciA....5.2307H,2022arXiv221109564P}) and possibly includes magnetic fields \citep{2022ApJ...931L..17K}, a more realistic, analytical theory of the interaction between convective turbulence and oscillations \citep{2022A&A...664A.164P}, or the combination of both.


\section*{Acknowledgments}

The authors are grateful to Karsten Brogaard and Mikkel Lund for reading and commenting on this manuscript. We thank also Tim Bedding, Hans-G\"{u}nter Ludwig and Luisa Rodr\'{i}guez D\'{i}az for valuable comments and fruitful discussions. 
RT is supported through NASA grants 80NSSC18K0559 and 80NSSC20K0543.
Funding for the Stellar Astrophysics Centre was provided by The Danish National Research Foundation (grant agreement no. DNRF106).
This research was supported by computational resources provided by the Australian Government through the National Computational Infrastructure (NCI) under the National Computational Merit Allocation Scheme and the ANU Merit Allocation Scheme (project y89). This work was supported by the Ministry of Education, Youth and Sports of the Czech Republic through the e-INFRA CZ (ID:90254).


%

\vspace{5mm}


\software{\texttt{Stagger} \citep{1995...Staggercodepaper,2018MNRAS.475.3369C,stein:StaggerCode}, \texttt{GARSTEC} \citep{2008Ap&SS.316...99W}, \texttt{ADIPLS} \citep{2008Ap&SS.316..113C}
}





\bibliography{References}{}
\bibliographystyle{aasjournal}



\end{document}